\documentclass{amsart}

\usepackage{mathtools}
\usepackage{amsmath}
\usepackage{amssymb}
\usepackage{amsfonts}
\usepackage{algorithmic}

\usepackage{color,subcaption}
\usepackage[pdftex]{graphicx}
\usepackage{bbm}
\usepackage{hyperref}
\usepackage{url}
\DeclareMathOperator*{\argmax}{arg\,max}

\hypersetup{hidelinks=true}
\usepackage{textcomp}
\def\BibTeX{{\rm B\kern-.05em{\sc i\kern-.025em b}\kern-.08em
    T\kern-.1667em\lower.7ex\hbox{E}\kern-.125emX}}
\begin{document}

\title[]{Unveil Sleep Spindles with Concentration of Frequency and Time}

\author{Riki Shimizu}
\address{Department of Biomedical Engineering, Duke University, Durham, NC, 27708 USA}
\email{riki.shimizu@duke.edu}

\author{Hau-Tieng Wu}
\address{Courant Institute of Mathematical Sciences, New York University, New York, NY, 10012 USA}
\email{hauwu@cims.nyu.edu}

\maketitle

\begin{abstract}
Objective: Sleep spindles contain crucial brain dynamics information. We introduce the novel non-linear time-frequency analysis tool 'Concentration of Frequency and Time' (ConceFT) to create an interpretable automated algorithm for sleep spindle annotation in EEG data and to measure spindle instantaneous frequencies (IFs).
Methods: ConceFT effectively reduces stochastic EEG influence, enhancing spindle visibility in the time-frequency representation. Our automated spindle detection algorithm, ConceFT-Spindle (ConceFT-S), is compared to A7 (non-deep learning) and SUMO (deep learning) using Dream and MASS benchmark databases. We also quantify spindle IF dynamics.
Results: ConceFT-S achieves F1 scores of 0.749 in Dream and 0.786 in MASS, which is equivalent to or surpass A7 and SUMO with statistical significance. We reveal that spindle IF is generally nonlinear.
Conclusion: ConceFT offers an accurate, interpretable EEG-based sleep spindle detection algorithm and enables spindle IF quantification. 
\end{abstract}

\section{Introduction}
{Sleep} spindles are brief bursts of activity within the sigma frequency range (approximately 11-16 Hz) of the electroencephalogram (EEG) signal, with durations ranging from 0.5 to 2 seconds, as noted by Iber et al. in 2007 \cite{iber2007aasm}. These spindles are physiologically generated by the thalamic reticular nucleus in coordination with specific thalamic nuclei, modulated by corticothalamic and thalamocortical connections, and manifest as spindle-shaped patterns in EEG readings. They are distinctive features of the N2 stage, a non-rapid eye movement (NREM) sleep phase signifying a transitional state between light and deep sleep. 

Understanding sleep spindles is crucial for unraveling the intricacies of sleep architecture, the processes governing sleep, memory consolidation, and broader cognitive functions \cite{hahn2019developmental}. These spindles also offer insights into healthy sleep patterns and neurological disorders.  For example, reduced sleep spindle activity and coherence have been observed in schizophrenia patients \cite{ferrarelli_reduced_2007,wamsley_reduced_2012}, suggesting impaired memory consolidation \cite{lustenberger_multidimensional_2015}. 
The grouping of spindle activity and fast brain oscillations by slow oscillations during slow-wave sleep represents an essential feature in the processing of memories during sleep \cite{molle_grouping_2002}.
Aging introduces changes in sleep spindles, potentially affecting their role in memory and sleep maintenance mechanisms \cite{martin_topography_2013}.
In epilepsy patients, the highest density of sleep spindles often appears away from the epileptic focus \cite{latka2005wavelet}, and alterations in sleep spindle density are linked to neurodegenerative disorders such as narcolepsy \cite{christensen2017sleep}. 
Notably, spindles exhibit variable oscillatory frequencies, characterized by ``instantaneous frequency'' (IF) fluctuations (alternative names include time-varying frequency and internal frequency modulation). Linear chirp rates for sleep spindle frequency acceleration or deceleration were quantified using matching pursuit with Gabor chirplet dictionaries \cite{schonwald2011quantifying}. IF dynamics were further explored with continuous wavelet transform  (CWT) \cite{zerouali2014time}, offering an alternative visualization in \cite[Figure 2]{zerouali2014time}. The absence of sleep spindle deceleration is linked to sleep-related disorders like sleep apnea \cite{carvalho2014loss}. More negative chirp rates are found in children with autism compared to normal controls \cite{tessier2015intelligence}. We refer readers to the state-of-the-art review of sleep spindle \cite{fernandez_sleep_2020} and its role in sleep disorders \cite{weiner2016spindle} for more details.

Conventional sleep spindle detection relies on manual EEG signal inspection, which is labor-intensive and prone to errors and inter-rater discrepancies among experts  \cite{devuyst2011automatic,stepnowsky2013scoring}. Consequently, there is a growing demand for reliable automated algorithms to streamline the process, reducing the burden on experts and improving spindle annotation consistency. Various algorithms have emerged over the years, encompassing time-frequency (TF) analysis \cite{lacourse_sleep_2019, huupponen_development_2007, yucelbas_automatic_2018}, 
matching pursuit \cite{schonwald_benchmarking_2006, larocco_spindler_2018}, signal decomposition \cite{parekh_detection_2015}, bayesian algorithms \cite{babadi_diba_2012}, decision trees \cite{duman_efficient_2009}, and support vector machines \cite{lachner-piza_single_2018}. 
Additionally, deep neural network (DNN) approaches have been introduced, such as the the DOSED algorithm based on the Convolutional Neural Network (CNN) framework \cite{chambon_dosed_2019}, the SpindleNet architecture that fuses a CNN and a Recurrent Neural Network (RNN) \cite{kulkarni_deep_2019} that has potential for real-time applications, and the U-Net architecture repurposed for spindle detection \cite{you_spindleu-net_2021, kaulen_advanced_2022}. Although DNN frameworks offer the potential to improve the precision of spindle detection, their inherent black-box nature restricts interpretability, and the demanding training phases required can pose practical application challenges.

This paper focuses on the application of a recently developed nonlinear TF analysis algorithm known as {\em Concentration of Frequency and Time} (ConceFT) \cite{DaWaWu2016} to spindle research. This application involves the creation of an automated annotation algorithm and an investigation into internal frequency modulation. ConceFT was designed to produce a precise and robust TFR for noisy time series data, improving the quantification of dynamics \cite{DaWaWu2016}. 
ConceFT consists of two components: a nonlinear version of the commonly used short-time Fourier transform (STFT) or CWT, known as the synchrosqueezing transform (SST) \cite{DaLuWu2011,Chen_Cheng_Wu:2014,sourisseau2022asymptotic}, and a nonlinear adaptation of the widely employed multitapering technique \cite{thomson_spectrum_1982,Percival:1993,babadi_review_2014}.
It is worth noting that the TFR generated by STFT or CWT tends to be blurred, even in noise-free signals, due to the uncertainty principle \cite{Ricaud2014}, limiting our ability to measure oscillatory component dynamics. SST addresses this issue by incorporating the phase information hidden in STFT or CWT. Additionally, the nonlinear generalization of multitapering reduces the impact of noise. In our context, where spindles are considered deterministic oscillatory components affected by the stochastic nature of EEG signals, ConceFT is a suitable tool for exploring spindles. For visual examples of TFR determined by STFT, CWT, and ConceFT, please refer to Figure \ref{Figure IF of spindles} and \cite[Figure 2]{zerouali2014time}.

\begin{figure*}[bht!]
\begin{center}
\includegraphics[trim=30 140 30 140,clip,width=1.2\textwidth]{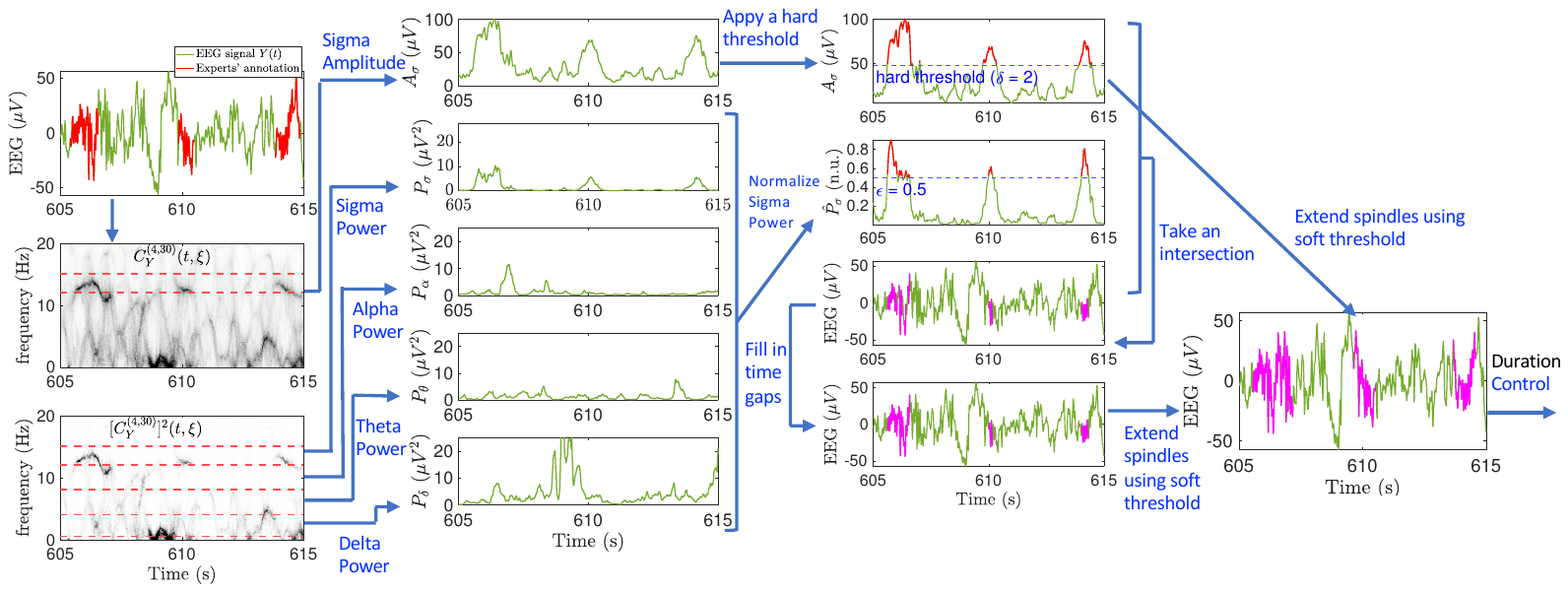}
\caption{\small A summary of the proposed automatic sleep spindle detection algorithm. First, apply ConceFT to produce the TFR of a raw EEG signal. Using the TFR, we find time indices where the normalized power over the sigma band [12-15] Hz exceeds the threshold determined by $\epsilon>0$. Separately, we find the time indices where the amplitude of the normalized sigma band exceeds the hard threshold set by $\delta$. Then, take the intersection of these two sets of indices. After filling the time gaps between neighboring detected events ($<$300 ms), we apply the soft threshold to the amplitude over the sigma band to extend the predicted spindles. Finally, duration criteria are imposed to remove spindles with a duration shorter than 300 ms or longer than 3,000 ms. A detailed description of the algorithm is given in Section \ref{subsection conceft-s algo}. \label{Figure overall flow}}
\end{center}
\end{figure*}

Our main contribution lies in leveraging ConceFT to develop an accurate and interpretable automated sleep spindle detection algorithm called {\em ConceFT-Spindle} (ConceFT-S).
In essence, we can perceive the Time-Frequency Representation (TFR) derived from ConceFT as a precise, time-varying power spectrum. This allows us to apply a bandpass filter concept to extract information relevant to sleep spindles. The concept behind ConceFT-S is straightforward: we calculate EEG energy within the sigma frequency band in the TFR generated by ConceFT and apply specific thresholding rules to identify the presence of spindles. ConceFT-S bears similarities to the approach described in \cite{combrisson_sleep_2017}, which utilizes CWT. For a comprehensive understanding of ConceFT-S, see Figure \ref{Figure overall flow}.

Our second contribution involves introducing a method to leverage ConceFT for enhanced quantification of spindle IF. In the existing literature, common approaches for studying spindle IF include matching pursuit \cite{schonwald2011quantifying} with linear chirplets and linear-type time-frequency analysis methods like STFT and CWT. 
It is worth noting that linear chirps and associated chirplets  have linear IF, making them inherently parametric methods. Consequently, unless the redundant frame used in the matching pursuit is expanded to encompass more general patterns, the estimated IF remains linear. However, as shown in Figure \ref{Figure IF of spindles}, spindle IF is generally nonlinear. This observation raises the question of whether it is possible to achieve a more accurate quantification of IF. While theoretically, expanding the redundant frame is feasible, to the best of our knowledge, it remains largely unexplored, and the associated computational complexity needs careful consideration. Given the nonparametric nature of spindle IF, an ideal tool is TF analysis, such as STFT and CWT, which are however limited by. We demonstrate that ConceFT offers a suitable approach to improve the quantification of spindle IF, revealing that spindle IF is typically nonlinear. In other words, the IF may not consistently accelerate or decelerate linearly over time.

The paper is structured as follows. Section \ref{section math background}provides the mathematical foundation, which includes a {\em phenomenological} model of sleep spindle, a concise overview of time-frequency analysis, and the fundamental principles of ConceFT along with its numerical implementation. Section \ref{section material and methods} offers a comprehensive introduction to the proposed automated sleep spindle detection algorithm, ConceFT-S, and the quantification of spindle IF. In Section \ref{section results}, we present the analysis results, demonstrating ConceFT-S' improved performance in spindle detection and an exploration of spindle IF using two benchmark databases. Sections \ref{section discussion} and \ref{section conclusion}, contain the discussion and conclusion of this paper, respectively.

\section{Mathematical background}\label{section math background}

We start with a {\em phenomenological} model of sleep spindle, followed by ConceFT and its numerical implementation. A quick summary of TF analysis is postponed to Section S-I in the Online Supplementary. 

\subsection{Mathematical model for spindles}

Sleep spindles are distinctive burst-like 10-15 Hz sinusoidal cycles observed in sleeping mammals' EEG. They are named after their spindle-like waveform \cite{fernandez_sleep_2020}. According to the American Association of Sleep Medicine (AASM), human sleep spindles manifest on the cortical surface as distinct waves with an 11-16 Hz frequency, typically within the 12-14 Hz range. These waves last more than 0.5 seconds and are most prominent in central derivations. Notably, spindle frequency can change, exhibiting speed acceleration or deceleration \cite{schonwald2011quantifying,carvalho2014loss}. See Figure \ref{Figure IF of spindles} for an illustration, which suggests that the acceleration of deceleration of spindle frequency is nonlinear. 
In short, sleep spindles are deterministic oscillations amid the stochastic EEG signal, combining defined characteristics with observed nonlinear frequency changes. In summary, sleep spindles contribute to the non-stationary nature of the EEG signal as follows:

\begin{figure}[hbt!]
\begin{center}
\includegraphics[width=0.85\textwidth]{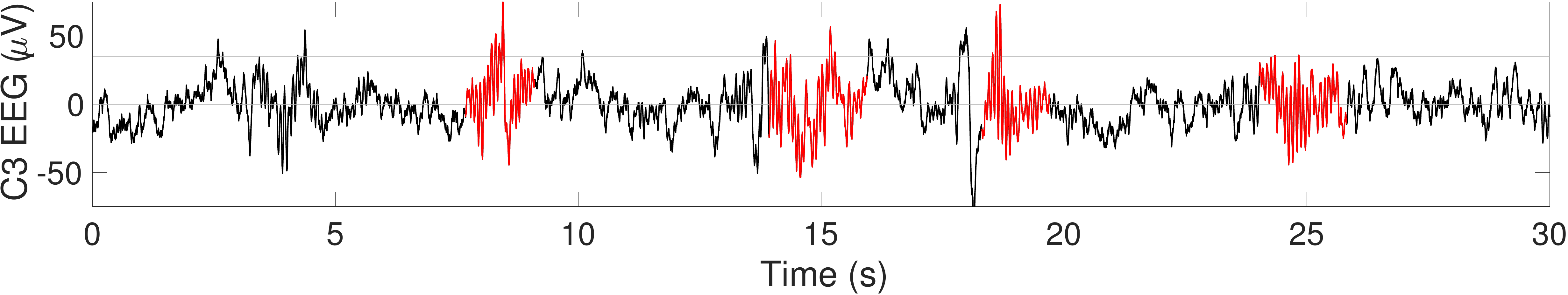}
\includegraphics[width=0.85\textwidth]{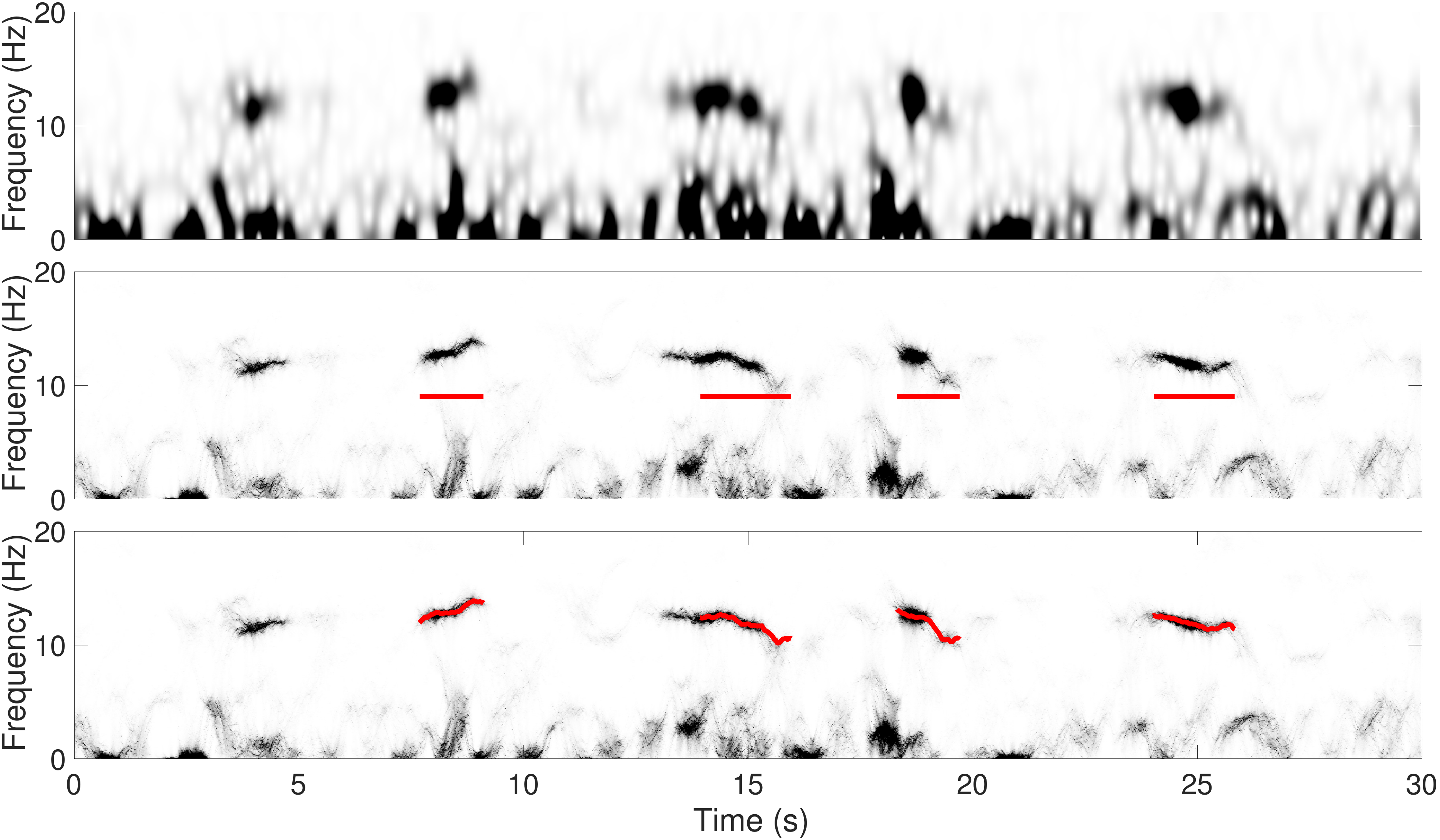}
\includegraphics[width=0.85\textwidth]{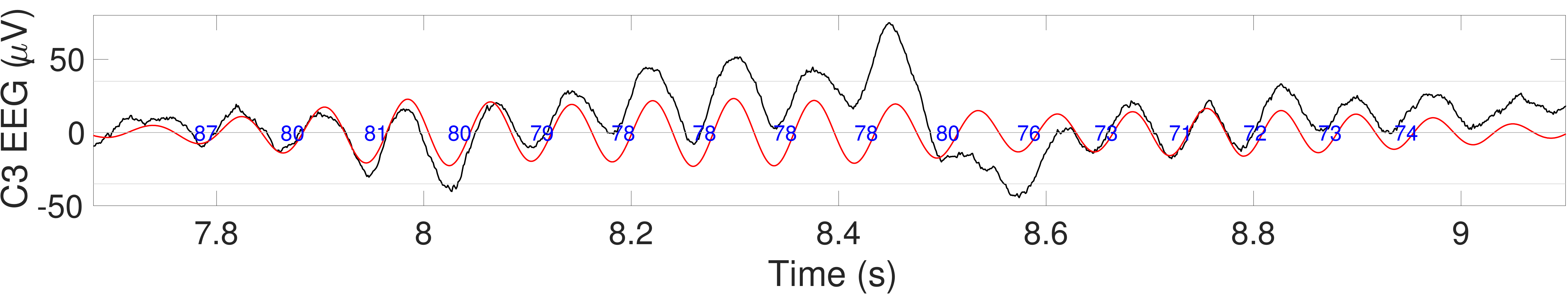}
\caption{\small An illustration of the time-varying frequency in a spindle. Top row: a typical 30 s EEG signal recorded during N2 sleep stage, where the sleep spindles labeled by experts are colored in red. Second row: the TFR of the EEG signal determined by STFT. 
Third and Fourth rows: the TFR of the EEG signal determined by ConceFT, where the sleep spindles labeled by experts are indicated by red lines in the third row, and the fitted instantaneous frequencies are superimposed as red curves in the fourth row.
Bottom: the zoomed in EEG segment of the first spindle labeled by experts is plotted in black, the band-passed signal over 10-15 Hz is superimposed in red, and the interval length of two consecutive cycles is shown in blue with the unit ms. \label{Figure IF of spindles}}
\end{center}
\end{figure}

\begin{enumerate}
\item[(F1)] The frequency and magnitude of spindle are time-varying. 

\item[(F2)] Spindles occurs in short-term.

\item[(F3)] Spindle coexists with the stochastic component of the EEG signal.
\end{enumerate}

In light of the notable characteristics (F1)-(F3), instead of directly modeling the complicated underlying mechanism of sleep spindles, we introduce a {\em phenomenological model} aimed at capturing the essential features of spindles within the EEG signal. We model an EEG signal with sleep spindles as
\begin{equation}\label{Model:equation}
E(t) = s(t)+\Phi(t),
\end{equation}
where $s(t)$ is a deterministic oscillatory signal describing the spindle and $\Phi$ is a random process describing the stochastic part of the EEG signal. The spindle $s$ fulfills the following assumptions. Fix a small constant $\epsilon>0$.
\begin{equation}\label{Model:equation2}
s(t) = \sum_{l=1}^L a_{l}(t)\cos(2\pi\phi_{l}(t))\,,
\end{equation}
where $L\in \mathbb{N}$, and for each $l=1,\ldots,L$,
\begin{itemize}
\item[(C1)] $a_{l}(t)>0$ is a bump-like $C^1$ function describing the envelope shape of the $l$-th spindle supported on an interval lasts for about 0.5-2 s, which we call the {\em amplitude modulation (AM)} ;  

\item[(C2)] $\phi_{l}(t)$ is a $C^2$ function that is strictly monotonically increasing denoting the {\em phase function} of the $l$-th spindle;

\item[(C3)] $\phi_{l}'(t)>0$ is the instantaneous frequency (IF) of the $l$-th spindle so that and $|\phi''_l(t)|\leq \epsilon \phi'_l(t)$ for all $t\in I_l$;
\end{itemize}
The density of spindle is quantified by $L$ divided by the epoch length under investigation. Over a 30 s epoch, $L$ could be roughly 6 in adults, 4 in elders, and more up to 20 in teenagers \cite{fernandez_sleep_2020}. 
Note that $L$, $\phi_l(t)$, $\phi_{l}'(t)$ and $a_{l}(t)$ might be different from one spindle to another, and depends on age, gender, genetics, recording location, etc \cite{hahn2019developmental,fernandez_sleep_2020}. For example, the support of $a_l$ is roughly 1-2 seconds in adults, a bit shorter in elders, and longer up to 4 s in infants. We do not assume any parametric form of $\phi'_l(t)$ besides constraining its time-varying speed in (C3), while in the literature, $\phi'_l$ is usually assumed to be linear \cite{schonwald2011quantifying}; that is, a linear chirp, or a cosine function in \cite{ktonas2009time}.
The stochastic part of the EEG signal $\Phi$ satisfies
\begin{itemize}
\item[(C4)] $\Phi(\cdot)$ is a long range dependent locally stationary random process and approximately stationary over a 30 s epoch. 
\end{itemize}
The scientific consensus holds that the EEG signal is non-stationary and possesses long-range temporal dependencies \cite{brismar2007human}. However, we can reasonably assume that, within a short time frame, the stochastic part of the EEG is ``relatively'' stationary. This leads us to model $\Phi$ as a locally stationary random process \cite{dahlhaus2012locally}, which imparts a form of stationarity over an epoch, typically lasting 30 seconds.
In summary, we depict the spindle as a deterministic oscillatory signal with preassigned frequency characteristics that coexists within the stochastic component of the EEG signal.

\subsection{A quick review of time-frequency analysis}\label{section summary TFA}

TF analysis tools can be broadly classified into three main categories: linear, bilinear, and nonlinear \cite{Flandrin:1999}. In general, these tools convert the input signal into a function defined on the TF domain or time-scale domain (or more generally time-frequency-chirp domain), called the TFR.
Linear-type TF analysis involves dividing the signal into segments and computing the spectrum for each segment. The choice of segmentation method distinguishes different approaches. For example, the STFT or Gabor transform employs a fixed window for segmentation, while the CWT adapts the segments based on wavelet dilation \cite{Flandrin:1999, Daubechies:1992}.
Bilinear-type TF analysis quantifies oscillatory properties using cross-correlation perspectives and various smoothing techniques, encompassing methods like the Wigner-Ville distribution and the Cohen class \cite{Flandrin:1999}.
Nonlinear-type TF analysis aims to represent signals in a data-driven manner, often modifying linear or bilinear TF analyses by incorporating phase information. Over recent decades, various practical methods have emerged, such as the reassignment method (RM), the SST, empirical mode decomposition (EMD), and several variations. For a comprehensive overview of the field, readers can refer to a recent review in \cite{wu2020current}. 
It is worth noting that the commonly employed matching pursuit method \cite{schonwald2011quantifying} can be viewed as a nonlinear-type TF analysis tool. In the matching pursuit technique, the signal is adaptively approximated using atoms within a pre-defined redundant frame. If needed, the signal can be subsequently transformed into a TFR by using the selected atoms.

In general, linear-type TF analysis methods like the STFT and CWT are susceptible to the uncertainty principle \cite{Ricaud2014}, which introduces blurriness in the resulting TFR. Additionally, they rely on the chosen window (or mother wavelet) and lack adaptivity to the signal characteristics.
Bilinear-type TF analysis, exemplified by the Wigner-Ville distribution, encounters limitations such as interference when dealing with signals composed of multiple oscillatory components or those with time-varying frequencies, even when a single oscillatory component is present \cite{Flandrin:1999}.
Nonlinear-type TF analysis emerges as a solution to these challenges. Among these methods, the widely used EMD lacks a theoretical foundation, potentially leading to erroneous interpretations with real data.
In contrast, the RM and SST have been rigorously developed with theoretical support \cite{DaLuWu2011,Chen_Cheng_Wu:2014,sourisseau2022asymptotic}. The SST and its variations incorporate phase information to mitigate blurriness caused by the uncertainty principle, resulting in a TFR less reliant on window choice \cite{DaLuWu2011,Wu:2011Thesis}. While the SST is a nonlinear method, its robustness to various types of noise, including non-stationary and heteroscedastic noises, has been established \cite{Chen_Cheng_Wu:2014}. However, its effectiveness can diminish when the signal-to-noise ratio (SNR) is low, typically below 1 dB. Thus, SST is a suitable tool when quantifying signal dynamics, like IF, is the focus and the SNR is not too low.

To enhance nonlinear-type TF analysis in low SNR conditions, one can consider the concept of {\em multi-tapering} (MT) \cite{Percival:1993}. 
The core idea behind MT involves using orthonormal windows, denoted as $h_1,\ldots,h_J$, to render noise components independent when employing methods like the RM or the SST with different windows. Averaging the results from RM or SST using these $J$ orthonormal windows helps mitigate the impact of noise \cite{Xiao_Flandrin:2007}. However, practical constraints limit the number of orthonormal windows due to the Nyquist rate, typically ranging from 6 to 10.
To overcome this limitation imposed by the Nyquist rate, the concept of {\em generalized MT} emerged, leading to the development of the ConceFT algorithm \cite{DaWaWu2016}. The fundamental idea behind the ConceFT is reducing the noise impact through generating more windows. Note that a point $\boldsymbol x:=(x_1,\ldots,x_J)\in \mathbb{C}^J$ on the unit sphere $\Omega^{J-1}\subset \mathbb{C}^J$ could lead to a linear combination of $J$ orthonormal windows, denoted as $h^{[\boldsymbol x]}:=\sum_{i=1}^Jx_ih_i$.
Consequently, using $n$ points on $\Omega^{J-1}$ results in $n$ TFRs via SST. Averaging these $n$ TFRs yields the desired TFR. Since the windows are not entirely independent, this approach deviates from the traditional MT technique. It's vital to note that the generalized MT technique leverages the nonlinearity of SST to break the dependence between windows effectively. This approach proves highly valuable in handling low SNR scenarios in real data \cite{Lin_Wu:2016,Wu_Liu:2018,Wu_Soliman:2018}, and its performance is guaranteed with theoretical underpinning \cite{DaWaWu2016}.

\subsection{SST and ConceFT in a nutshell}

We start with a summary of SST. Suppose the signal we want to analyze is a realization of a generalized random process $Y$ with $\mathbb{E}Y=f$ and a finite covariance structure, where $f$ could be as general as a tempered distribution. Take a window function $h(t)$ that is real symmetric centered at $0$, smooth and decay fast so that $h$ is a Schwartz function. We assume that $h$ is the Gaussian function to simplify the discussion. Mathematically, the STFT of $Y$ associated with $h$ is defined as 
\begin{equation}\label{Definition:STFT}
V_Y^{(h)}(t,\xi):=\langle Y,\, h(\cdot-t)e^{-i2\pi \xi (\cdot-t)}\rangle\,,
\end{equation}
where $t\in\mathbb{R}$ is the time, $\xi \in \mathbb{R}^+$ is the frequency, $h$ is the window function chosen by the user, and the notation $\langle\cdot,\cdot\rangle$ means the evaluation of the random process $Y$ with the Schwartz function $h(\cdot-t)e^{-i2\pi \xi (\cdot-t)}$, which can be formally understood as $\int Y(x)h(x-t)e^{-i2\pi \xi (x-t)}\mathrm{d} x$. In other words, STFT is obtained by dividing the signal into pieces, and evaluate the Fourier transform of each piece. By patching them together, we obtain the information about how the signal oscillates at each time.  
Second, evaluate the {\em reassignment rule}, 
\begin{equation}
\omega^{(h)}_{f}(t,\xi):=
\nu - \mathfrak{Im}\frac{V_f^{(\mathcal{D}h)}(t,\xi)}{2\pi V_f^{(h)}(t,\xi)},
\label{eq:reass}
\end{equation}
where $\mathfrak{Im}$ means taking the imaginary part, and $\mathcal{D}h$ is the derivative of $h$ \cite[Definition 2.3.12]{Wu:2011Thesis}. Equation~(\ref{eq:reass}) is well-defined on every point $(t,\nu)$ where $V_f^{(h)}(t,\nu)\neq 0$.
Third, the linear TFR determined by STFT is sharpened by:
\begin{equation}\label{Definition:1stSST}
S^{(h)}_{f}(t,\nu):=
\int_{\mathfrak{N}_t}V^{(h)}_f(t,\xi)\delta_{|\nu-\omega^{(h)}_f(t,\xi)|}\mathrm{d} \xi,
\end{equation}
where $\mathfrak{N}_t:=\{\xi:\,|V^{(h)}_f(t,\xi)|>\theta_0\}$ \cite[Definition 2.3.13]{Wu:2011Thesis} and $\theta_0>0$ is the chosen threshold. 
The main step to sharpen the linear TFR is using the {\em phase} information of STFT via \eqref{eq:reass}. 
Hence, \eqref{Definition:1stSST} should be understood as two separate steps:
\begin{itemize}
\item Select all entries $(t,\xi)$ so that the frequency information provided by $\omega^{(h)}_f(t,\xi)$ is $\nu$  via $\delta_{|\nu-\omega^{(h)}_f(t,\xi)|}$;
\item Gather all non-zero STFT coefficients to the entry $(t,\nu)$. 
\end{itemize}
The SST can be effectively applied to investigate the adaptive harmonic model \cite[Theorem 2.3.14]{Wu:2011Thesis}; that is, a multicomponent oscillatory signal with each component oscillating with slowly varying amplitude and frequency. For instance, in the case of $f(t)=A(t)\cos(2\pi\phi(t))$, the TFR determined by SST concentrates around $\phi_l'(t)$ while encoding the AM function $A(t)$ as the intensity. This representation is notably less influenced by the choice of window and robust the noise \cite{DaLuWu2011,Chen_Cheng_Wu:2014,sourisseau2022asymptotic}. Although not necessary for our current discussion, it is worth noting that SST allows users to achieve more signal processing missions, e.g., recovering individual oscillatory components or denoising.

With SST, we now describe ConceFT. 
Take $J$ orthonormal windows, $h_1,h_2,\ldots,h_J\in L^2(\mathbb{R})$, where $J\in \mathbb{N}$. We focus on the first $J$ Hermite windows due to its property of having a minimal essential support in the TF domain \cite{DaWaWu2016}. For $\mathbf{x}:=(x_1,\dots,x_J)\in \Omega^{J-1}$, we have a new window $h_{\mathbf{x}}:= \sum_{j=1}^{J} x_j h_j$, which satisfies $\|h_{\mathbf{x}}\|_{L^2}=1$. 
Fix $Q\in \mathbb{N}$ much larger than $J$ and randomly uniformly sample $Q$ points from $\Omega^{J-1}$, denoted as $\mathbf{x}_1,\ldots,\mathbf{x}_Q$. The ConceFT of $Y$ is 
\begin{equation}\label{eq:conceft_quadrature}
	C_Y^{(J,Q)}(t,\nu)
	:=\frac{1}{Q}\sum_{k=1}^{Q} \left|S^{(h_{\mathbf{x}_k})}_Y(t,\nu)\right|\,,
\end{equation} 
where $t\in \mathbb{R}$ and $\nu>0$. The traditional MT approach \cite{Xiao_Flandrin:2007} is a special ConceFT when $Q=J$, and $\mathbf{x}_k=e_k$, the unit vector with the $k$-th entry $1$. Since more than $J$ non-orthonormal windows are used in ConceFT, it is called the ``generalized MT'' scheme. Since $\mathbf{x}_k$ and $\mathbf{x}_i$ are in general not orthonormal, the noises in $V^{(h_{\mathbf{x}_k})}_{f}(t,\nu)$ and $V^{(h_{\mathbf{x}_i})}_{f}(t,\nu)$ are dependent.
However, the nonlinearity of SST drives the noise in $S^{(h_{\mathbf{x}_k})}_{f}(t,\nu)$ and $S^{(h_{\mathbf{x}_i})}_{f}(t,\nu)$ to be less correlated compared with that in $V^{(h_{\mathbf{x}_k})}_{f}(t,\nu)$ and $V^{(h_{\mathbf{x}_i})}_{f}(t,\nu)$ \cite{DaWaWu2016}. Due to the reduced correlation, the impact of noise on the final TFR is reduced via the averaging. In practice, ConceFT is especially effective when the SNR is low. 
See Figure \ref{Figure IF of spindles} for a comparison of TFRs determined by STFT and CWT.  In STFT, four labeled spindles produce blurred representations due to the uncertainty principle, limiting IF quantification. ConceFT sharpens these bumps into curves, resulting in a cleaner TFR with reduced influence from EEG background noise. Furthermore, ConceFT significantly concentrates the spectral spreading in the low-frequency region (around 0-3Hz) compared to STFT, particularly around slow oscillations. This enables us to extract IF from the TFR using a curve fitting algorithm, as explained in the next section.

\subsection{Numerical implementation}\label{numerical implementation of CONCEFT}
Denote the discretized signal as $\mathbf{f} \in \mathbb{R}^N$ at the sampling rate $f_s>0$, where $N\in \mathbb{N}$. So the recording duration is $T = N/f_s$ s. Assume the recording starts at time $0$.  
To evaluate the ConceFT of $\mathbf{f}$, we take the top $J\in \mathbb{N}$ hermite windows and uniformly sample them over the interval $[-10, 10]$ at a sampling period of $dt = \frac{10}{K}$. Denote the discretized Hermite windows as $\mathbf{h}_1, \mathbf{h}_2,  \mathbf{h}_3 \in \mathbb{R}^{2K+1}$. Also denote $\mathbf{h}'_1,\mathbf{h}'_2,\mathbf{h}'_3 \in \mathbb{R}^{2K+1}$ as the derivatives of these Hermite windows.  We then uniformly sample $Q$ random points $z_1, \ldots, z_{Q} \in \Omega^{J-1} \subset \mathbb{C}^J$ and obtain $Q\in \mathbb{N}$ new window functions $\mathbf{g}_1, \ldots, \mathbf{g}_{Q} \in \mathbb{C}^{2K + 1}$ (and their derivatives $\mathbf{g}'_1, \ldots, \mathbf{g}'_{Q} \in \mathbb{C}^{2K + 1}$) by the formulas
$\mathbf{g}_i = \sum_{j=1}^J z_i(j) \mathbf{h}_j$ and $\mathbf{g}'_i = \sum_{j=1}^J z_i(j) \mathbf{h}'_j$, where $i = 1, \ldots, Q$. 
For each $i = 1, \ldots, Q$, the STFT of $\mathbf{f}$ with the window $\mathbf{g}_i$, denoted as $\mathbf{V}_i \in \mathbb{C}^{N \times M}$, is evaluated by
$\mathbf{V}_i(n, m) = \sum_{k=1}^{2K+1} \mathbf{f}(n + k - K - 1) \mathbf{g}_i(k) e^{\frac{-i2\pi (k-1)m }{M}}$,
where $M\in \mathbb{N}$ is the number of frequency bins in the frequency axis, $m=1,\ldots,M$, $n=1,\ldots,N$, and we set $\mathbf{f}(l) := 0$ when $l < 1$ or $l > N$. Here, $M$ can be arbitrarily picked, which balanced between the frequency resolution and the computational time. Similarly, we define the other STFT of $\mathbf{f}$ using $\mathbf{g}'_i$, denoted as $\mathbf{V}'_i \in \mathbb{C}^{N \times M}$.
To sharpen each $\mathbf{V}_i$, we choose a threshold $\upsilon > 0$ and calculate the {\em reassignment operators}
$\mathbf{\Omega}_i(n, m) = 
-\mathfrak{Im}\frac{N}{2\pi}\frac{\mathbf{V}_i'(n,m)}{\mathbf{V}_i(n,m)}$ 
when $|\mathbf{V}_i(n,m)|> \upsilon$ and $
-\infty$  when $|\mathbf{V}_i(n,m)|\leq \upsilon$,
where $i = 1,\ldots,Q$. 
With the reassignment operator, the SST of $\mathbf{f}$ with the window $\mathbf{g}_i$, denoted as $\mathbf{S}_i\in \mathbb{C}^{N \times M}$, is evaluated by 
$\mathbf{S}_i(n, m) = \sum_{l;\, |l- \mathbf{\Omega}_{\mathbf{i}}(n, m)|<\nu} \mathbf{V}_i(n, l)$,
where $\nu>0$ is a small constant. 
Finally, the TFR of $\mathbf{f}$ determined by ConceFT, denoted as $\mathsf{CFT}_\mathbf{f} \in \mathbb{C}^{N \times M }$, is 
\begin{gather}
\mathsf{CFT}_\mathbf{f}(n, m) = \frac{1}{Q}\sum_{i=1}^{Q} |\mathbf{S}_i(n, m)|\,.
\end{gather}

\section{Materials and Methods}\label{section material and methods}

\subsection{Annotated Databases} \label{subsection material}
We consider two open-access benchmark databases in this study. 
The first dataset is the {\em Dream} database from the University of MONS-TCTS Laboratory and Universite Libre de Bruxelles-CHU de Charleroi Sleep Laboratory \cite{devuyst_dreams_2005}, which was previously used to evaluate automatic spindle detection algorithms \cite{you_spindleu-net_2021, parekh_detection_2015, kulkarni_deep_2019, jiang_robust_2021}. It contains 30 minutes of Polypolysomnography (PSG) recordings from 8 subjects with various sleep disorders with a sampling frequency at 50 Hz for one subject, 100 Hz for another subject, and 200 Hz for the other subjects. The dataset was annotated by two experts. Expert 1 annotated all 8 subjects, while Expert 2 annotated the first 6 subjects. Subjects 1 and 3 were annotated over the C3-A1 channel, while other subjects were annotated over the CZ-A1 channel. The original EEG signals were resampled into 50 Hz for the purpose of standardization. For the first 6 subjects, which were annotated by both experts, the union of the two scorings was used as the ground truth. The intra-rater agreement was reported in \cite[Table 1]{liu_evaluating_2017}. Denote $e_1, e_2\in \{0,1\}^N$ to represent experts' annotations, where $0$ indicates no spindles and $1$ indicates the existence of spindle. The union annotation is determined by an element-wise OR operation on $e_1$ and $e_2$. The last 2 subjects were not used in this research following the conventions of previous research \cite{you_spindleu-net_2021, jiang_robust_2021}.

The second database is the second subset (SS2) of the Montreal Archive of Sleep Studies (MASS) \cite{oreilly_montreal_2014}. The use of this dataset was approved by the Duke Institutional Review Board. This subset comprises full-night PSG recordings of 19 young and healthy participants. Two experts separately annotated sleep spindles on the EEG channel C3-A1. Expert E1 annotated all 19 recordings, while Expert E2 made annotations for 15 of them. The EEG signals were sampled at 256 Hz but were resampled at 50 Hz for the standardization purpose in this research. It is worth mentioning that while Expert E1 adhered to the standard AASM scoring guidelines, Expert E2 used a similar approach to Ray et al. \cite{ray_validating_2010} that utilized both broad-band EEG signals (0.35-35 Hz) and sigma-filtered signals (11-17 Hz). 
Furthermore, Expert E2 did not set a minimum duration for spindles, and four of the 19 nights were not assessed due to perceived poor sleep quality or inconsistent signal integrity. The significant difference in the annotated spindles by E1 and E2 is well known \cite[Table 2]{liu_evaluating_2017}, so we used the same union scheme in the {\em Dream} database as the ground truth, and only 15 subjects were utilized, following a common practice in the literature \cite{kulkarni_deep_2019, jiang_robust_2021}.

\subsection{Proposed ConceFT-S Algorithm}\label{subsection conceft-s algo}

The overall flowchart of the proposed ConceFT-S algorithm is shown in Figure \ref{Figure overall flow}, where the EEG signal from Subject 2 in the Dream Dataset between 605 to 615 seconds is demonstrated. 
Below we detail the algorithm step by step, and provide details about parameter selection.

\subsubsection{Step 1: Evaluate ConceFT}

Denote the raw EEG signal as $\mathbf{f} \in \mathbb{R}^N$ sampled at the sampling rate $f_s>0$. Compute ConceFT of $\mathbf{f}$ with the first $J$ Hermite windows, where the Hermite window implementation is detailed in Section \ref{numerical implementation of CONCEFT}. Denote the result as $\mathsf{CFT}_\mathbf{f} \in \mathbb{C}^{N \times M }$. 

\subsubsection{Step 2: Indices for Spindle Detection} 

Given $\mathsf{CFT}_\mathbf{f}$, we calculate the {\em sigma band amplitude} from 12 Hz to 15 Hz. These values picked to account for the spectral leakage outside of the 12-14 Hz, the range used by Combrisson et al \cite{combrisson_sleep_2017}. 

\begin{equation}
    \mathsf{A}_{\sigma}(n) =\Delta_f \sum_{m: 12 \leq \mathsf{freq}(m) \leq 15} \mathsf{CFT}_{\mathbf{f}}(n, m)\,,
\end{equation}
where $\mathsf{freq}(m)$ is frequency associated with the $m$-th frequency index and $\Delta_f:=\mathsf{freq}(2)-\mathsf{freq}(1)$. 
Next, we calculate the {\em power density} of four frequency bands, where the delta band is $B_\delta:=[0.5-4]$ Hz, the theta band is $B_\delta:=[4-8]$ Hz, the alpha band is $B_\alpha:=[8-12]$ Hz, and the sigma band is $B_\sigma:=[12-15]$ Hz, by 
\begin{align}
    \mathsf{P}_{\texttt{Band}}(n) &= \frac{\Delta_f }{|B_{\texttt{Band}}|} \sum_{m:  \mathsf{freq}(m) \in B_{\texttt{Band}}} |\mathsf{CFT}_{\mathbf{f}}(n, m)|^2\,,
\end{align}
where $\texttt{Band}=\delta,\theta,\alpha,\sigma$.
The normalized sigma band power is then computed by dividing $\mathsf{P}_{\sigma}(n)$ by the sum of the power in the four bands
\begin{align*}
    \hat{\mathsf{P}}_{\sigma}(n) = \frac{\mathsf{P}_{\sigma}(n)}{\mathsf{P}_{\delta}(n) + \mathsf{P}_{\theta}(n) + \mathsf{P}_{\alpha}(n) + \mathsf{P}_{\sigma}(n)}\,.
\end{align*}

Introduce a threshold parameter $\delta>0$ such that the hard threshold for the sigma band amplitude is $\theta_a:=$mean$(\mathsf{A}_{\sigma}) + \delta \times $std$(\mathsf{A}_{\sigma})$. We apply this hard threshold to $\mathsf{A}_{\sigma}$ and get a subset of indices $T_1 \subset \{1, \ldots N\}$; that is,  $i \in T_1$ if $\mathsf{A}_{\sigma}(i)$ is greater than the hard threshold. At the same time, we apply the second threshold parameter $\epsilon$ to $\hat{\mathsf{P}}_{\sigma}(n)$ and get another subset of indices $T_2 \subset \{1, \ldots N\}$; that is, $i \in T_2$ if $\hat{\mathsf{P}}_{\sigma}(i)$ is greater than $\epsilon$. We do not use the mean and standard deviation for $\hat{\mathsf{P}}_{\sigma}(n)$ because it is already normalized to take into account the individual signal characteristics. 
Define $\mathcal{I}=T_1\cup T_2$. 

\subsubsection{Step 3: Post-processing}
Cluster $\mathcal I$ into $\mathcal I_1,\ldots, \mathcal I_L$ so that each $\mathcal I_i$ contains consecutive integers, $\mathcal I_k\cap \mathcal I_l=\emptyset$, $\mathcal I=\cup_{l=1}^L\mathcal I_l$, and indices in $\mathcal I_i$ are smaller than those in $\mathcal I_{i+1}$. Due to the randomness of the EEG signal, we cannot directly use $\mathcal I_1,\ldots, \mathcal I_L$ to estimate sleep spindles, and we modify them by the following rules before estimating sleep spindles. 
First, if the distance between $\mathcal I_i$ and $\mathcal I_{i+1}$, defined as $\texttt{dist}(\mathcal I_i,\,\mathcal I_{i+1})=\min_{k\in \mathcal I_i,\,l\in\mathcal I_{i+1}}\{|k-l|\}$, is less than 300 ms for any $i$, bridge the time gap by replacing $\mathcal I_i$ by $\mathcal I_i\cup \mathcal I_{i+1}$ and deleting $\mathcal I_{i+1}$. Suppose we end up with a new set of intervals, $\mathcal I_1,\ldots, \mathcal I_{L'}$, where $L'\leq L$.
Subsequently, define a soft threshold calculated from the sigma band amplitude by $\vartheta_a:=$0.5$\times$(mean$(\mathsf{A}_{\sigma}) + \delta \times $std$(\mathsf{A}_{\sigma})$). 
For each $1\leq i\leq L'$, denote $\mathcal I_i:=\{s_i,s_i+1,\ldots,e_i\}$. Suppose $s_i'$ and $e_i'$ are the closest points at which the sigma band amplitude intersects with $\vartheta_a$. Update $\mathcal I_i$ by $\mathcal I'_i:=\{s'_i,s'_i+1,\ldots,e'_i\}$. Denote the new set of intervals as $\mathcal I'_1,\ldots, \mathcal I'_{L'}$.
Once more, if the distance between $\mathcal I'_i$ and $\mathcal I'_{i+1}$ is less than 300 ms for any $i=1,\ldots,L'-1$, bridge the time gap by replacing $\mathcal I'_i$ by $\mathcal I'_i\cup \mathcal I'_{i+1}$ and deleting $\mathcal I'_{i+1}$. Suppose we end up with a new set of intervals, $\mathcal I'_1,\ldots, \mathcal I'_{L''}$, where $L''\leq L'$.
Finally, any $\mathcal I'_i$ shorter than 300 ms or longer than 3,000 ms are discarded, and we end up with the final set of intervals, $\mathcal I'_1,\ldots, \mathcal I'_{L'''}$, where $L'''\leq L''$, which are our final spindle estimates. The post-processing steps follow Combrison et al \cite{combrisson_sleep_2017} with a slight modification in the duration criterion. The cutoff durations of 500 ms and 2,000 ms were used in \cite{combrisson_sleep_2017} while we changed them to 300 ms and 3,000 to fit the characteristics of the datasets used in this research. Their code implementations are available at \url{https://github.com/EtienneCmb/visbrain}.

\subsubsection{Parameter selection}

In Step 1, choose $J=3$, $K = f_s$, $Q=30$ and $M = 4000$ over the frequency range $[0,20]$ Hz in our implementation of ConceFT. The choice of $K$ is based on the rule of thumb in the TF analysis that the chosen window should encompass approximately $10$ cycles of the oscillatory component, or more if the signal is noisy. Considering that our target oscillations are in the sigma band ($11-16$ Hz), $K = f_s$ corresponds to a one-second signal span, encompassing approximately $11$ to $16$ oscillations. This range effectively mitigates the influence of the stochastic EEG component. While theoretically, a higher $Q$ might enhance performance, empirical evidence shows that performance plateaus around $20$ or $30$ \cite{DaWaWu2016}. Given the trade-off between performance and computational efficiency, we set $Q=30$.
For further details, see the Supplemental Material in \cite{DaWaWu2016}.

In Step 2, the two threshold parameters are optimized through a non-exhaustive grid search in the training phase, where we search $\delta$ from $\{1,1.5,2,2.5,3\}$ and $\epsilon$ from $\{0.05,0.2,0.35,0.5\}$. This optimization is chosen to balance between the prediction accuracy and computational speed.

\subsection{Spindle IF estimate}\label{section spindle if est}

With the TFR determined by ConceFT, $\mathsf{CFT}_\mathbf{f} \in \mathbb{C}^{N \times M }$, and the detected spindle or spindle labeled by experts, we could estimate the spindle IF by fitting a curve into the TFR by solving the following optimization problem.
\begin{equation}
c^*=\argmax_{c:\{1,\ldots,n\}\rightarrow \mathcal M}
\sum_{\ell=1}^n
\mathbf{R}(\ell,c(\ell))
-
\lambda\sum_{\ell=1}^{n-1}|\Delta c(\ell)|^2\,,
    \label{singleCurveExt}
\end{equation}
where the spindle is assumed without loss of generality to live in the first $n$ samples to simplify the discussion, $\mathcal M\subset \{1,\ldots,M\}$ is the frequency band $[10,15]$ Hz, $\Delta c:\mathcal{M}_-^{n-1}$, where $\mathcal{M}_-:=\{i-j|\, i,j\in \mathcal{M}\}$ so that $\Delta c(\ell):=c(\ell+1)-c(\ell)$ for $\ell=1,\ldots,n-1$, $\lambda>0$ is the penalty term constraining the regularity of the fit curve $c$, and $\mathbf{R}(\ell,q)=\log\frac{|\mathsf{CFT}_\mathbf{f}(\ell,q)|}{\sum_{i=1}^N\sum_{j=1}^M|\mathsf{CFT}_\mathbf{f}(i,j)|}$. Here, $\Delta c$ is the numerical differentiation of the curve $c$ and $\mathbf{R}(\ell,q)$ is a normalization of the TFR. This optimization can be efficiently solved by the penalized forward-backward greedy algorithm \cite{Chen_Cheng_Wu:2014}. 

To make a connection with existing literature that often assumes that the spindle IF is linear, we carry out the following curve fitting scheme. Suppose the spindle IF is $c^*$ over interval $[-t_c,\,t_c]$ extracted by \eqref{singleCurveExt}. Fit a quadratic polynomial $\tilde{c}(t)=\beta_{0}+\beta_{1} t+\frac{1}{2}\beta_{2}t^2$ into $c^*$ by minimizing the least squared error, where $\beta_{0}$, $\beta_{1}$ and $\beta_{2}$ are the mean rate, linear chirp rate and quadratic chirp rate with the unit Hz, Hz$/s$ and Hz$/s^2$ respectively, and calculate the relative root mean square error (RRMSE), which is defined as $\|\tilde{c}-c^*\|_2/\|c^*\|_2$. For a comparison, fit a linear polynomial $\check{c}(t)=\gamma_{0}+\gamma_{1} t$ into $c^*$ by minimizing the least squared error, where $\gamma_{0}$ and $\gamma_{1}$ are the mean rate and linear chirp rate with the unit Hz and Hz$/s$ respectively, and calculate the RRMSE. We assume spindle IFs are independent, so $(\beta_0,\beta_1,\beta_2)^\top$ (and $(\gamma_0,\gamma_1)^\top$) of different spindles are independent.

\subsection{Comparison with existing detection algorithms}
We compare the proposed ConceFT-S algorithm with two state-of-the-art automatic spindle detection algorithms, A7 \cite{lacourse_sleep_2019} and SUMO \cite{kaulen_advanced_2022}. A7 has the best performance among non-DNN-based algorithms, and SUMO has the best performance among DNN-based algorithms.

\subsubsection{A7}
The A7 algorithm operates by setting thresholds on four key parameters: both the absolute and relative power within the sigma band (11-16 Hz), as well as the covariance and correlation between broadband-filtered EEG signals (0.3-30 Hz) and signals filtered within the sigma band (11-16 Hz) \cite{lacourse_sleep_2019}. 
The authos showed that the A7 algorithm achieved the best F1 score on the younger cohort and the second best on the older cohort. It is shown in \cite{you_spindleu-net_2021} that A7 outperforms DETOKS \cite{parekh_detection_2015}. Thus we focus on A7 in this work and use the A7 implementation in \url{https://github.com/swarby/A7_LacourseSpindleDetector}.

\subsubsection{SUMO}
The SUMO algorithm is a one-dimensional variant of the U-Net architecture tailored for sleep spindle detection. Kaulen et al. benchmarked SUMO using the MODA dataset, demonstrating its superior performance over A7 in sensitivity, recall, and F1 score \cite{kaulen_advanced_2022}. This algorithm is thus chosen in this study. Spindle U-Net \cite{you_spindleu-net_2021} achieved a similarly good performance but was not used in this research as it shares the same architecture as SUMO. To apply SUMO to the Dream dataset, we adjust the training approach to suit the data's specifics. We divided the 1800 s EEG signals from each individual into fifteen 120-s segments. This 120 s segment length was chosen because it is the nearest integer divisor of 1800 to 115, the segment length employed in \cite{kaulen_advanced_2022}. For every test participant, the 75 blocks from the other 5 subjects were allocated to training and validation datasets in a 13:2 split. The implementation of SUMO is available at \url{https://github.com/dslaborg/sumo}.

\subsection{Statistical Analysis}

All quantities are presented as the mean $\pm$ standard deviation. Continuous variables are analyzed using the Wilcoxon ranksum test. P values $<0.05$ are considered statistically significant with Bonferroni correction. 

We applied a leave-one-subject-out cross-validation (LOSOCV) scheme to assess ConceFT-S performance on both datasets. In each fold, we selected $\delta$ and $\epsilon$ by finding the combination that maximizes the average F1 score on the training subjects. These thresholds were then used to evaluate ConceFT-S on the test subject. We reported average performance metrics across all train-test splits for each dataset. We adopted the analysis-by-event approach \cite{warby_sleep_2014}to reliably identify sleep spindles, comparing estimated spindles to the ground truth, as described in Section \ref{subsection material} on an event basis. 
For each spindle in the expert annotation, we calculated the temporal intersection with the closest estimated spindle divided by their union. If this relative overlap exceeded a threshold (set to $0.2$, a common suggestion in the literature \cite{jiang_robust_2021, lacourse_sleep_2019}), we counted it as a true positive (TP). If an expert-annotated spindle lacked any detected spindle with sufficient relative overlap, we considered it a false negative (FN). Similarly, if a detected spindle lacked any annotated spindle with sufficient relative overlap, we categorized it as a false positive (FP). We reported sensitivity (SEN), precision (PRE), and F1 score (F1) as performance metrics.

\section{Results}\label{section results}

The computation was conducted in the MATLAB R2022a environment using a 2GHz Quad-Core Intel Core i5 processor and 16GB of RAM. The Matlab implementation of the proposed algorithm can be found in \url{https://github.com/rsbci/ConceFT-Spindle}. The sensitivity analysis of ConceFT-S can be found in Section S-II in the Online Supplementary.

\subsection{Data Visualization}
\begin{figure}[hbt!]
\begin{center}
\includegraphics[trim=5 40 0 40, clip,width=0.95\textwidth]{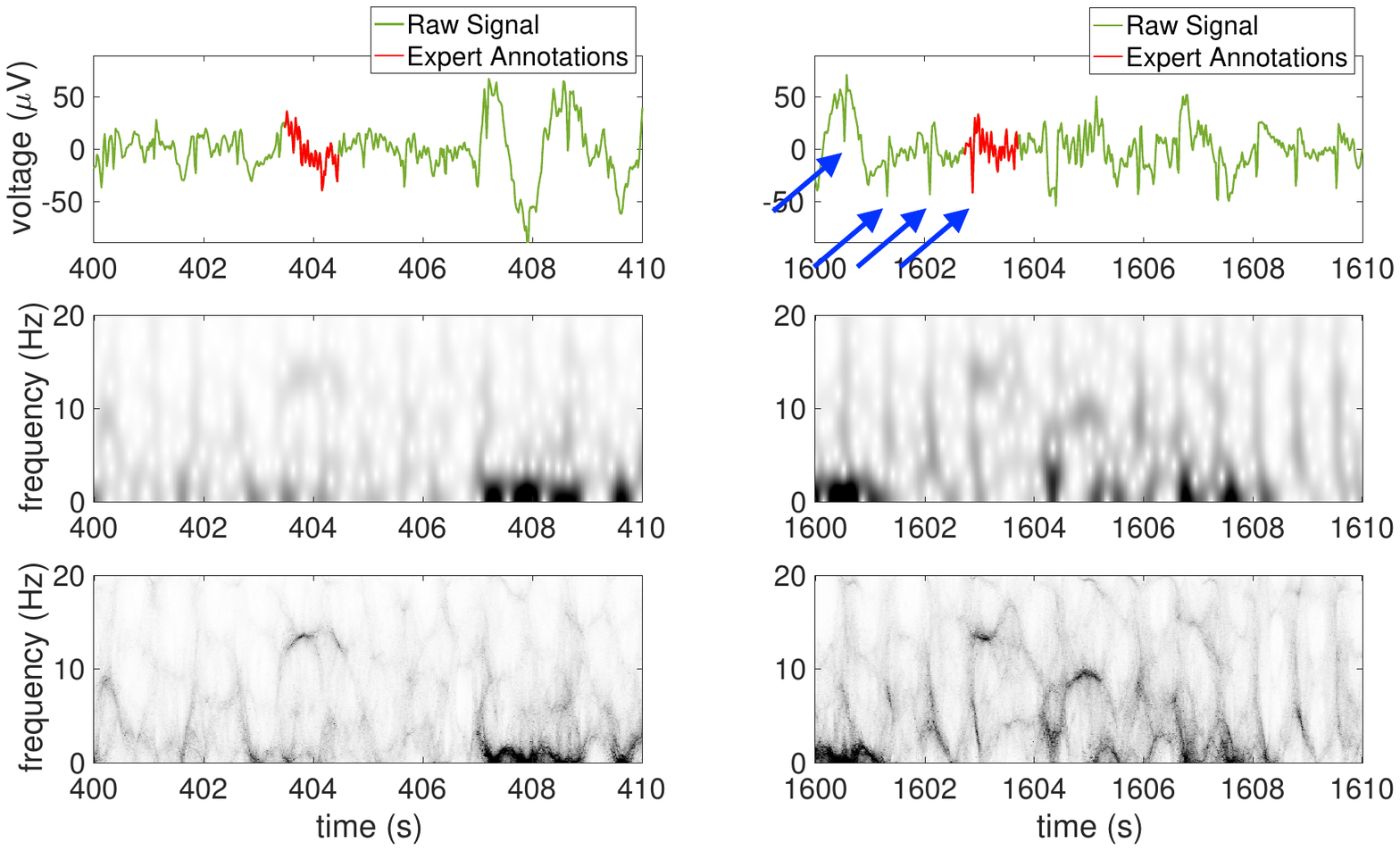}
\caption{\small An illustration of TFRs determined by STFT (middle row) and ConceFT (bottom row). The EEG signal on the top left (right respectively) subplot is from 400 to 410 seconds (1600 to 1610 seconds respectively) is from Subject 2 in the Dream Dataset. \label{Figure TFR example}}
\end{center}
\end{figure}

We start by visually assessing ConceFT's effectiveness. Figure \ref{Figure TFR example} shows a comparison of different TFRs given by STFT and ConceFT, utilizing segments from Subject 2 in the Dream dataset. 
In the first segment, the experts annotated a spindle around the 404th s while another spindle around the 1603th s was annotated for the second segment. In both TFRs, we can observe energy concentration in the sigma band for each spindle. 
However, ConceFT exhibits sharper power concentration compared to STFT, especially around the 404th second. Additionally, ConceFT significantly reduces spectral spreading in the low-frequency region of STFT (0-3Hz) around the 408th second and 1600th second. Another notable observation is the cardiogenic artifact (indicated by blue arrows). While this artifact is relatively prominent in STFT, ConceFT mitigates its impact.

\subsection{Performance of ConceFT-S}

In Tables~\ref{Table1}, we show the SEN, PRE, and F1 values for the three detection algorithms tested, including A7, SUMO and ConceFT-S. The ConceFT-S achieves an average sensitivity of 0.709, precision of 0.807, and F1 score of 0.749 on the Dream Dataset, and an average sensitivity of 0.789, precision of 0.801, and F1 score of 0.786 on the MASS SS2 subset. ConceFT-S outperforms A7 in F1 with statistical significance, where the $p$-value is $p < 10^{-4}$  on MASS and $p=0.0411$ in Dream. On the other hand, ConceFT-S outperforms SUMO in F1 without statistical significance in both datasets, where the $p$-value is $p = 0.132$ in Dream and $p = 0.836$ on MASS. The thresholds used for testing were stable across subjects within each dataset, with $\delta$ equal to 2 and $\epsilon$ to 0.5 for the Dream dataset and $\delta$ equal to 2.5 and $\epsilon$ to 0.2 for the MASS dataset.

\begin{table}[bht!]
\renewcommand{\arraystretch}{1.3}
\caption{\small LOSOCV results of by-event evaluation of different sleep spindle detection algorithms.}
\label{Table1}
\centering
\begin{tabular}{ccccc}
\hline
Database & Algorithm & SEN &  PRE & F1 \\
\hline
 & A7 \cite{lacourse_sleep_2019} & $0.586$ &  $\textbf{0.818}$ & $0.678$ \\
Dream & SUMO \cite{kaulen_advanced_2022}& $0.634$ &  $0.773$ & $0.674$ \\
 & ConceFT-S & $\textbf{0.709}$ &  $0.807$ & $\textbf{0.749}$ \\
\hline
 & A7 \cite{lacourse_sleep_2019} & $0.619$ &  $0.806$ & $0.692$ \\
MASS & SUMO \cite{kaulen_advanced_2022}& $0.758$ &  $\textbf{0.828}$ & 0.782 \\
 & ConceFT-S & $\textbf{0.789}$ &  $0.801$ & $\textbf{0.786}$ \\
\hline
\end{tabular}
\end{table}

Next, we consider the summary of average spindle density and duration in the N2 sleep stage for each subject in the MASS and Dream datasets, by experts' annotation and different algorithms. 
The calculation of spindle duration was conducted on spindles that were labeled TP and reported in median $\pm$ median absolute deviation since there exists 6 statistical outliers (longer than 5 seconds) in the spindle duration in the MASS dataset, while the spindle density is reported in mean $\pm$ standard deviation. The spindle durations (densities respectively) of experts' annotation were 1.14$\pm$0.43 s (0.104$\pm$0.03 per second respectively) for the MASS dataset and 1.0$\pm$0.10 s (0.065$\pm$0.025 per second respectively) for the Dream dataset. On average the spindle length is longer and the density is higher in MASS compared with Dream, which probably comes from the annotation strategies or the existence of subjects with sleep disorders in Dream. 
Meanwhile, the spindle durations (densities respectively) determined by ConceFT-S were 0.96$\pm$0.32 s (0.101$\pm$0.002 per second respectively) for the MASS dataset and 1.0$\pm$0.34 s (0.057$\pm$0.024 per second respectively) for the Dream dataset. Except the spindle durations in the MASS dataset ($p < 10^{-4}$), there is no significant difference between expert annotations and predictions by ConceFT-S. 
The RRMSE of spindle durations (densities respectively) between annotations and predictions by A7, SUMO and ConceFT-S are 
$0.213$, $0.085$ and $ 0.232$ respectively ($0.330$, $0.353$ and $0.202$ respectively) in the Dream dataset, and
$0.220$, $0.131$ and $ 0.186$ respectively ($0.275$, $0.234$ and $0.238$ respectively) in the MASS dataset.
Note that the while overall ConceFT-S does not outperform SUMO, the performance is comparable.

\subsection{Exploration of spindle instantaneous frequency}

To explore dynamics of spindles, see Figure \ref{Figure IF of spindles} for an illustration of a 30 s EEG signal with several spindles. It is evident that the spindle cycles' durations are not constant; they decrease, as quantified by the IF condition (C2) in the phenomenological model \eqref{Model:equation}. This time-varying frequency is visualized in the TFR of the EEG signal determined by conceFT in the middle panel in Figure \ref{Figure IF of spindles}. A closer look at the dominant curve in the TFR and the fitted curve shows that the curve is not linear. That is, the spindles shown in Figure \ref{Figure IF of spindles} are not linear chirps.

\begin{figure}[t!]
\begin{center}
\begin{subfigure}{0.9\textwidth}
        \includegraphics[width=1\textwidth]{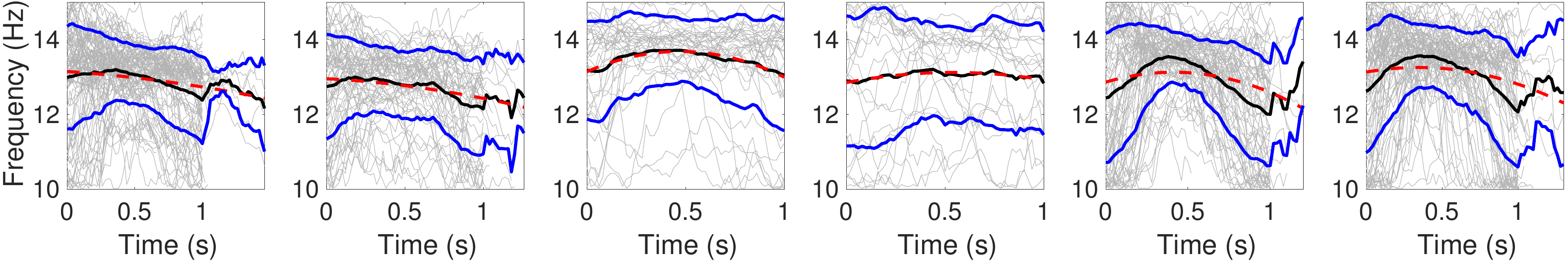}
          \caption{Dream dataset}
          \label{fig:NiceImage1}
      \end{subfigure}
      
\begin{subfigure}{0.9\textwidth}
\centering
        \includegraphics[width=0.835\textwidth]{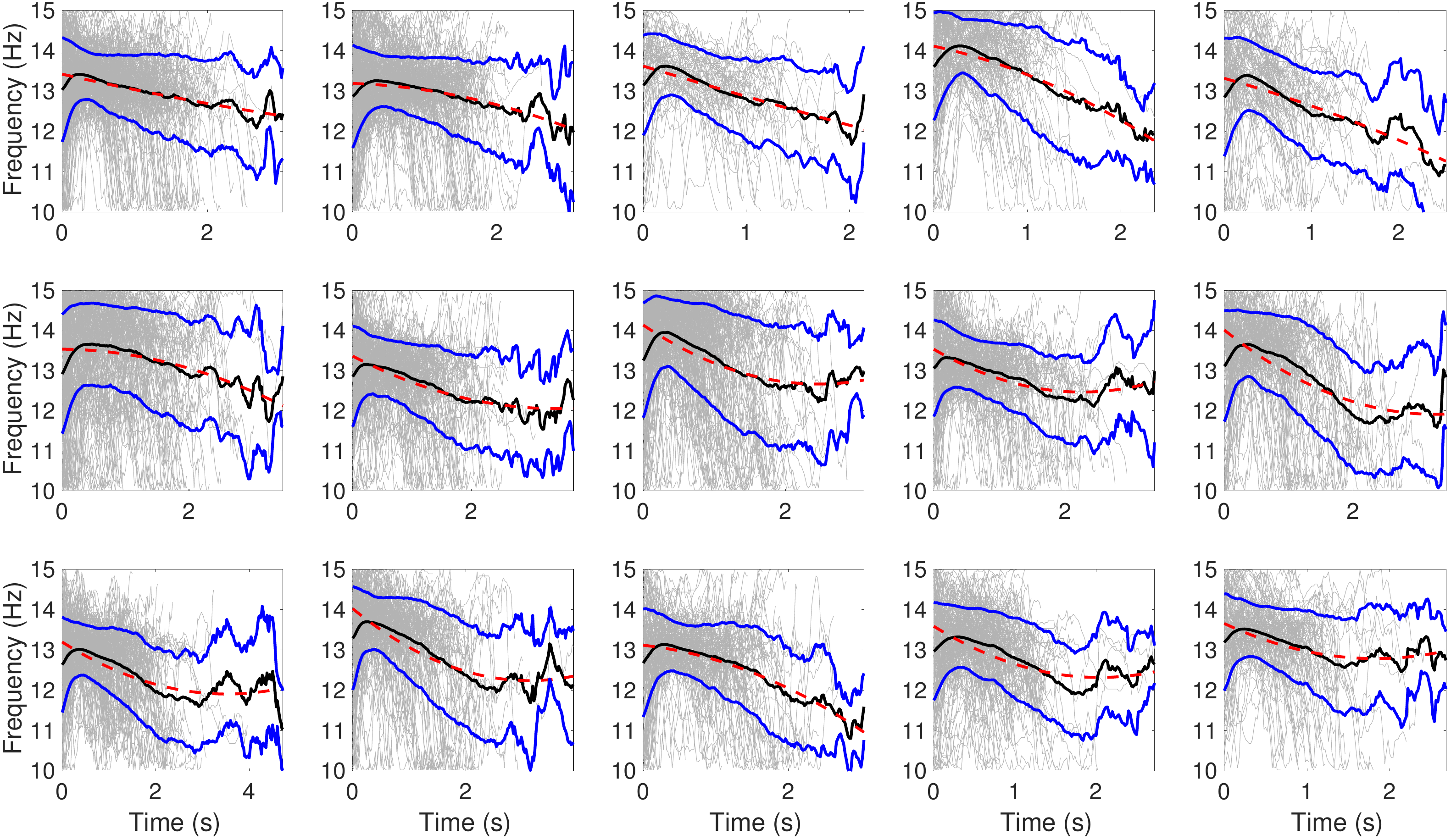}
          \caption{MASS dataset}
          \label{fig:NiceImage2}
      \end{subfigure}
\caption{\small The top row is from the Dream dataset, and the remaining three rows are from the MASS dataset. Each subplot shows the functional plot of all spindles (in gray) of one subject, and the associated mean and mean $\pm$ standard deviation of IFs are superimposed as black and blue curves respectively. The red dashed curve is the fitted quadratic polynomial to the mean IF. \label{Fig instantaneous frequency}}
\end{center}
\end{figure}

To further explore the spindle dynamics in terms of IF, we first explore IF on the subject level. For each subject in each dataset, gather the IFs of all spindles labeled by experts using the curve extraction \eqref{singleCurveExt}. We align all spindles by their associated middle points of the labels, denote $c_{i,j}(t)$ to be the estimated IF of the $j$th spindle of the $i$-th subject on $[-t_{i,j},t_{i,j}]$, where $t_{i,j}>0$. Then we assess the mean and standard deviation of all IFs at each time $t$ when there are at least $3$ spindles that last longer than $t$. See Figure \ref{Fig instantaneous frequency} for an illustration. In each subject we show a functional plot of all spindles (in gray) of one subject, along with the associated mean and mean $\pm$ standard deviation of IFs. It is clear that the averaged IF is not linear.

\begin{figure}[t!]
\begin{center}
\includegraphics[width=0.3\textwidth]{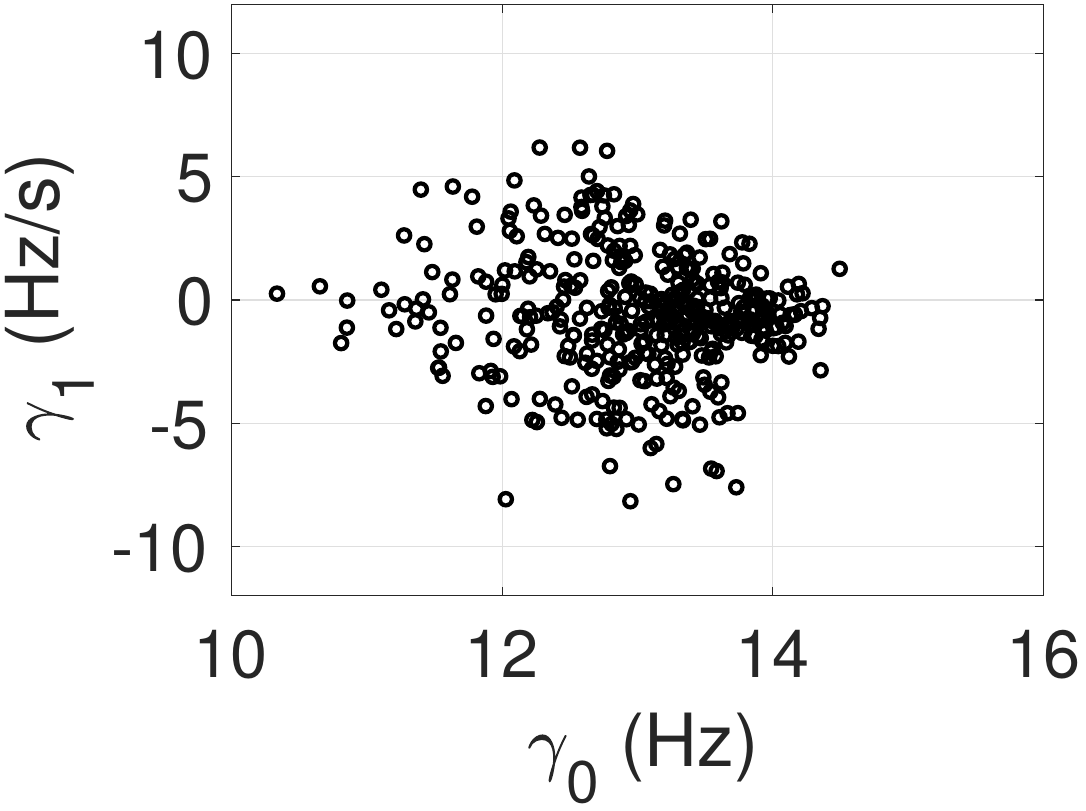}
\includegraphics[width=0.3\textwidth]{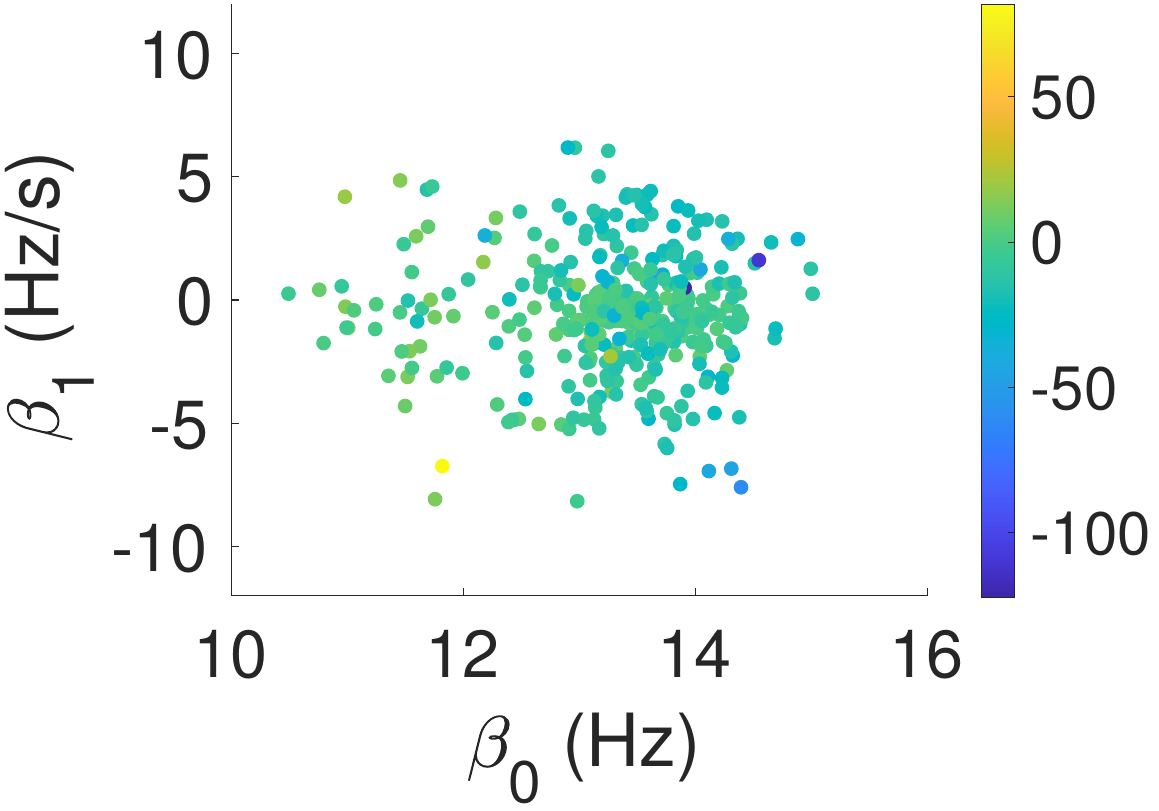}
\includegraphics[width=0.3\textwidth]{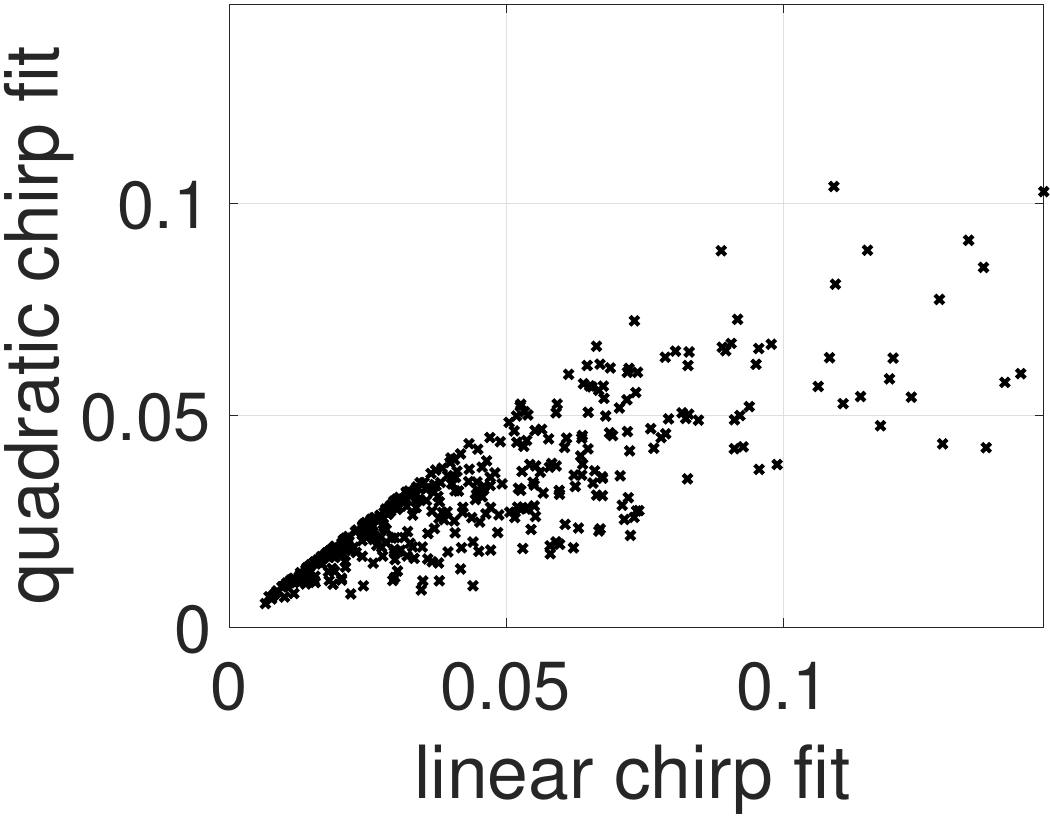}\\

\includegraphics[width=0.3\textwidth]{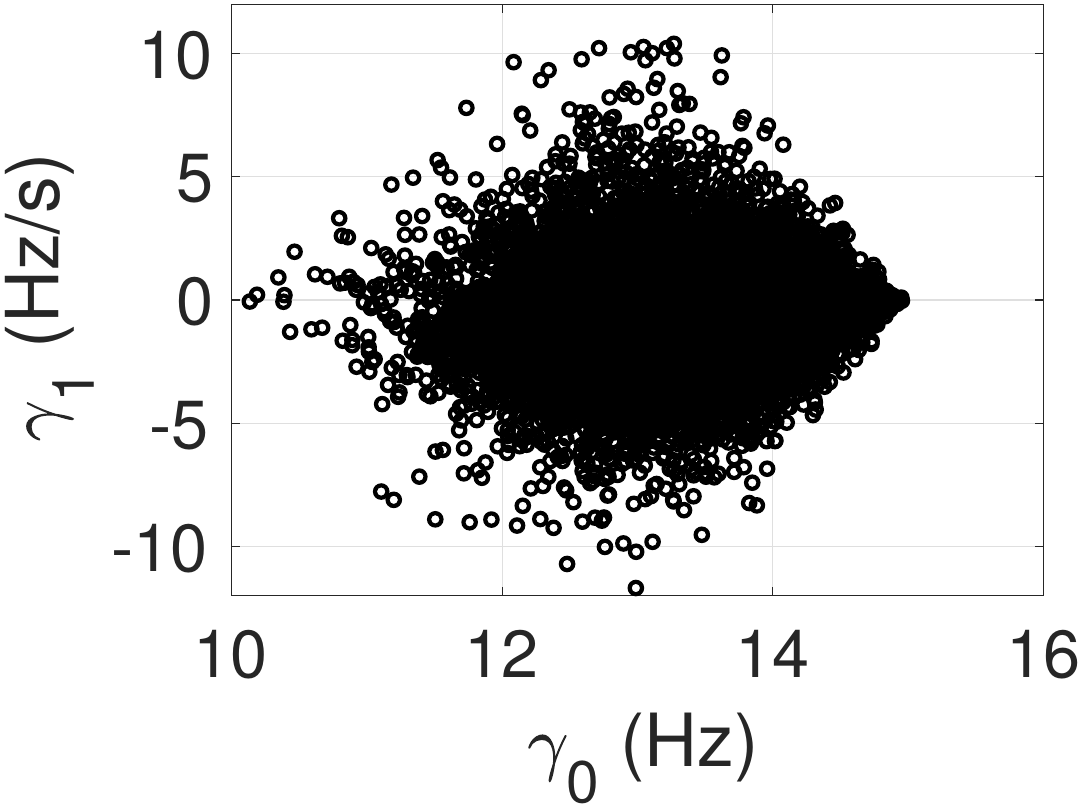}
\includegraphics[width=0.3\textwidth]{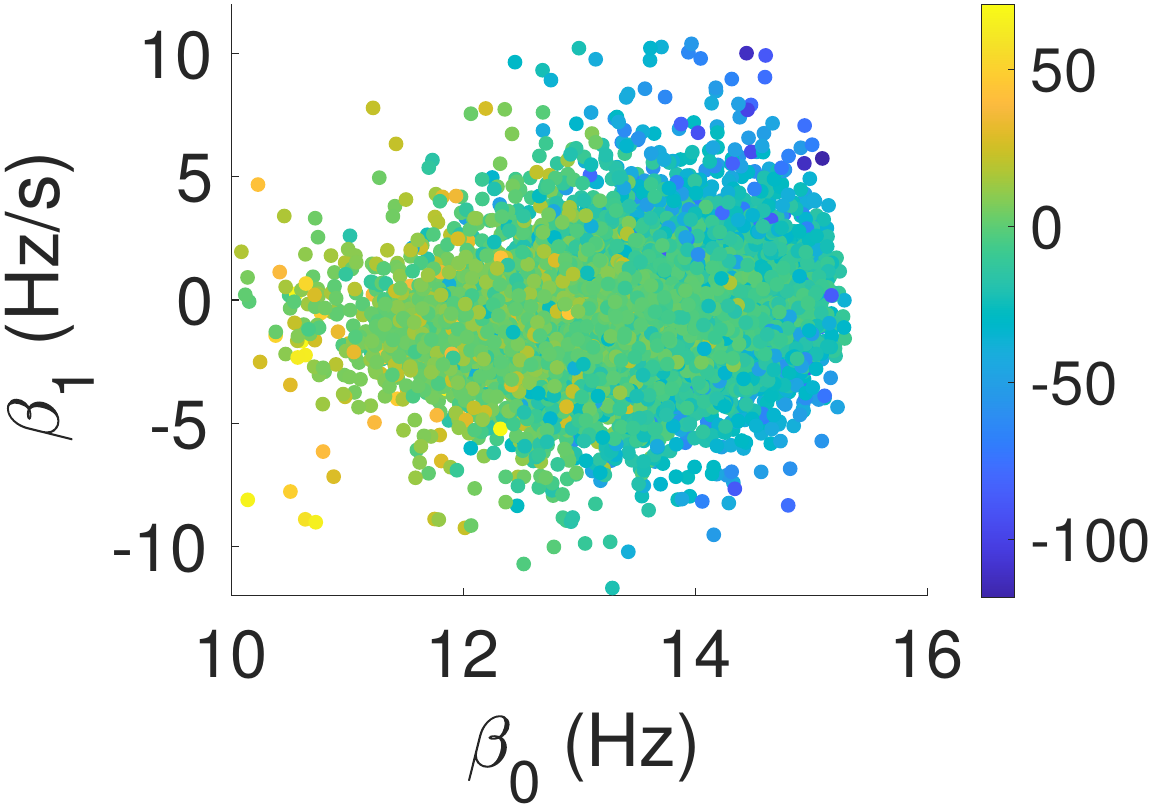}
\includegraphics[width=0.3\textwidth]{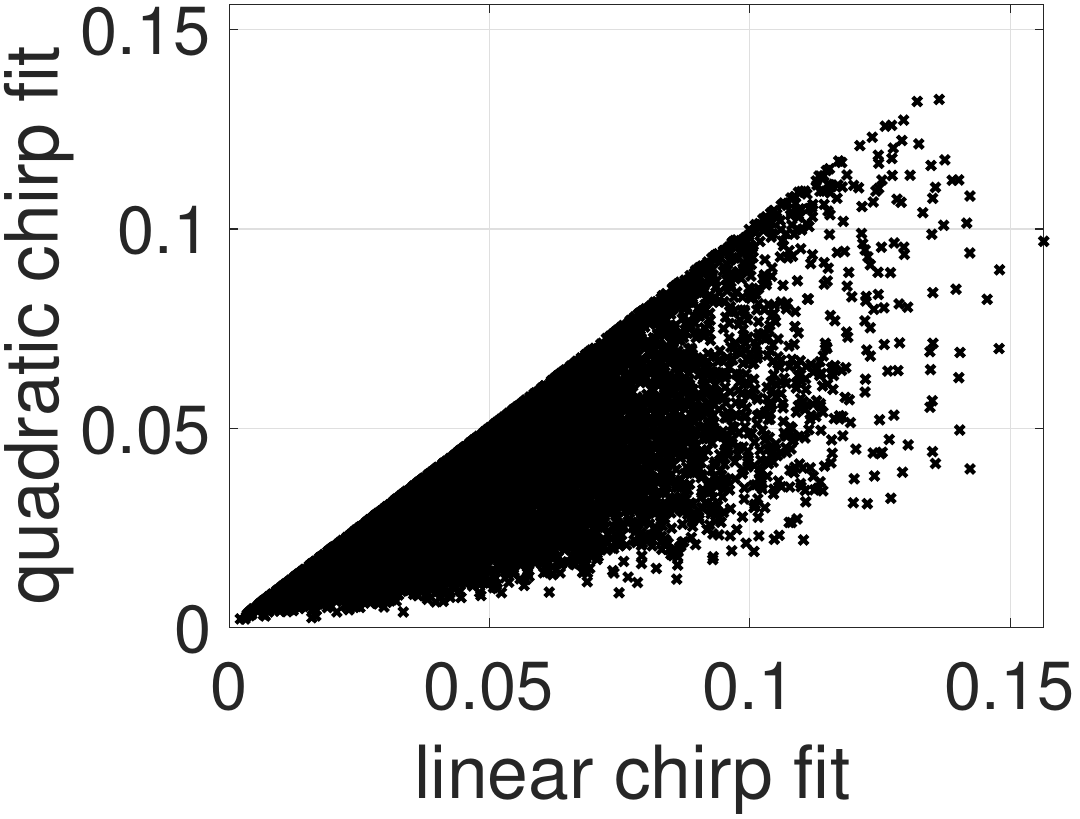}
\caption{\small An exploration of spindle IF on the spindle level. Left column: the fitting of spindle IF by the linear chirp, where the x-axis is the mean rate $\beta_0$, and the y-axis is the linear chirp rate $\beta_1$. Middle column: the fitting of spindle IF by the quadratic chirp, where the x-axis is $\beta_0$, the y-axis is $\beta_1$, and the color indicates the quadratic chirp rate $\beta_2$. Right column: the scatterplot of the relative RMSE using the linear chirp fit against the relative RMSE using the quadratic chirp fit. The top row is the result from the Dream dataset, and the bottom row is from the MASS dataset. \label{Fig instantaneous frequency2}}
\end{center}
\end{figure}

Next, we quantify the IF on the spindle level. In the Dream (MASS respectively) dataset, 63.8\% (65.3\% respectively) of spindles have negative linear chirp rate, and 71.0\% (71.3\% respectively) of spindles have negative quadratic chirp rate, which both have significant difference with $p<10^{-8}$ ($p<10^{-8}$ respectively) by applying the binomial test with the null hypothesis that the positive and negative rates are of the same ratio. 
The distributions of $(\beta_0,\beta_1,\beta_2)^\top$, $(\gamma_0,\gamma_1)^\top$ and RRMSE are shown in Figure \ref{Fig instantaneous frequency2}, where the mean $\pm$ standard deviation of $\beta_0$, $\beta_1$, $\beta_2$, $\gamma_0$ and $\gamma_1$ in the Dream (MASS respectively) dataset are $13.313\pm  0.797$,  $-0.589\pm 2.414$, $-7.924\pm 14.8695$, $13.009\pm 0.735$ and $-0.731\pm  2.382$ respectively ($13.427\pm 0.706$, $-0.526\pm 1.724$, $-4.902\pm 11.841$, $13.207\pm 0.627$ and $-0.599\pm 1.715$ respectively). In the Dream (MASS respectively) dataset, the Pearson's linear correlation coefficients between $\beta_1$ and $\beta_2$, $\beta_1$ and $\beta_3$ and $\beta_2$ and $\beta_3$ are $-0.01$, $-0.355$ and $-0.125$ respectively, where only $\beta_1$ and $\beta_3$ and $\beta_2$ and $\beta_3$ are different from 0 with statistical significance with $p<10^{-8}$ and $p=0.01$ respecitvely ($0.114$, $-0.393$ and $-0.097$ respectively, where all are different from 0 with statistical significance and $p<10^{-8}$).
The RRMSEs of the Dream and MASS datasets are $0.034\pm 0.018$ and $0.025 \pm 0.013$ respectively.
In both datasets, the quadratic chirp fits better than the linear chirp with statistical significance, where the $p<10^{-8}$ by applying the Wilcoxon signed-rank test to the RRMSEs of linear and quadratic polynomial fits. This result supports that the spindle IF is in general not linear.

\subsection{Sensitivity analysis of ConceFT-S}

We test the robustness of ConceFT-Spindle algorithm with respect to its parameters, mainly $M$, $J$, and $Q$ as a sensitivity analysis. The first parameter of interest is $M$, which indicates the frequency resolution of TFR in the frequency domain. For $M=40,400,4000$, the F1 scores are 0.698, 0.717, and 0.749 respectively for the Dream dataset, and the F1 scores are 0.773, 0.782, and 0.786 respectively for the MASS dataset. On the other hand, for each corresponding $M$, the computational times are 4, 10, and 45 seconds for calculating the ConceFT of a 30-second EEG signal. This shows that there is no significant decline in the F1 score as the $M$ decreases. The $p$-value between $M=400$ and $M=4000$ is $p=0.329$ on Dream and $p = 0.619$ on MASS while $p$-value between $M=40$ and $M=4000$ is $p=0.310$ on Dream and $p = 0.340$ on MASS. Thus. when a fast computation is needed, we could speed up the computation by slightly sacrificing the performance.

\begin{figure}[hbt!]
\begin{center}
\includegraphics[width=0.5\textwidth]{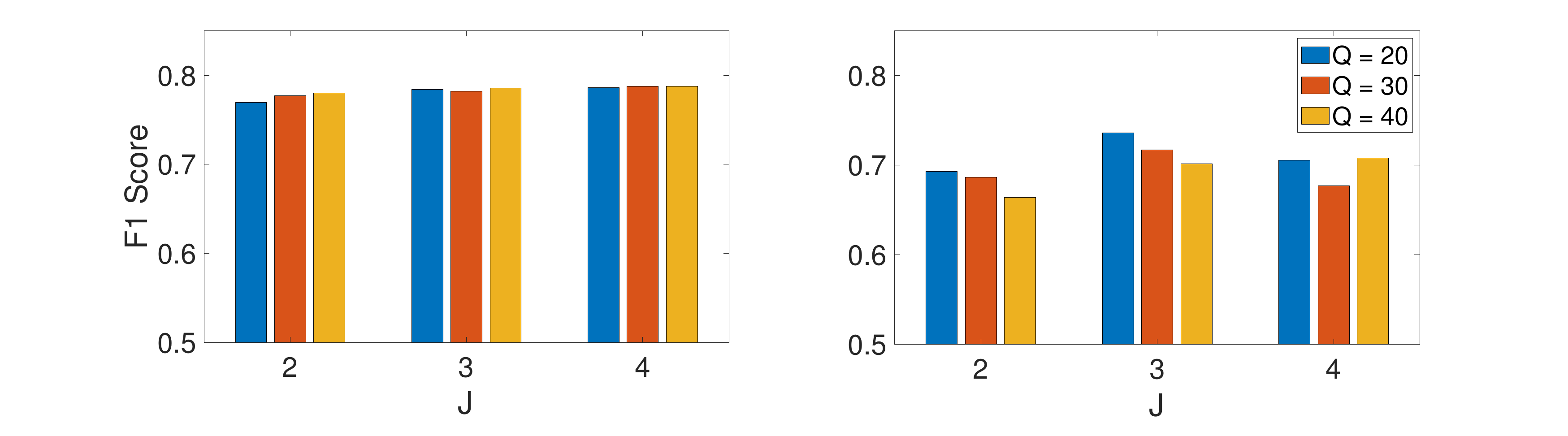}
\caption{\small Sensitivity analysis of ConceFT-S, where we show the F1 score of ConceFT-S with respect to $J=2, 3,4$ and $Q=20, 30, 40$. The left plot is from the MASS dataset, and the right plot is from the Dream dataset. \label{Sensitivity}}
\end{center}
\end{figure}

Figure \ref{Sensitivity} shows the F1 value of ConceFT-S with $J=2,3,4$ and $Q=20, 30, 40$. To speed up the calculation, this analysis was conducted with $M=400$. Across both datasets, the F1 score is the lowest when $J=2$ while there is no obvious trend between $J=3$ and $J=4$. Changing $Q$ does not seem to have a consistent effect on the F1 score. Overall, the performance of the algorithm is stable across datasets and parameters.

\section{Discussion}\label{section discussion}

In this work, we introduce a novel Time-Frequency (TF) analysis tool known as ConceFT for the study of sleep spindles. Our contributions can be summarized as follows. The first contribution is developing an automatic spindle detection algorithm, ConceFT-Spindle (ConceFT-S), based on ConceFT, and compare it with two state-of-the-art detectors, A7 \cite{lacourse_sleep_2019} and SUMO \cite{kaulen_advanced_2022}, on two public benchmark databases. 
Our second contribution focuses on demonstrating how ConceFT can be applied to investigate the spindle IF. We present evidence that the spindle IF exhibits non-linearity, challenging the assumption of a linear chirp frequency across time.

The overall performance of ConceFT-S is equivalent to, if not better than, existing results in the literature. Notably, You et al. \cite{you_spindleu-net_2021} and Jiang et al. \cite{jiang_robust_2021} have assessed an extensive range of spindle detectors on the same dataset, utilizing the identical Leave-One-Subject-Out Cross-Validation (LOSO-CV) methodology applied in our research. Their studies reveal a wide spectrum of F1 scores, ranging from 0.175 for DETOKS to 0.739 for Spindle U-Net, with Spindle U-Net sharing structural similarities with SUMO and both being rooted in the U-Net neural network framework. While deep learning techniques have consistently displayed superior performance over traditional methods in sleep spindle detection, our findings underscore that straightforward and interpretable thresholding techniques, when combined with an improved TFR, can yield comparable results to deep learning methods. Notably, compared with ``black-box'' deep learning models, an advantage of ConceFT-S is its interpretability, which is essential for scientific research. Grounded in the amplitude and power within the sigma frequency band, it offers transparent insights into the detection process, fostering understanding and trust, and bridging the gap between computational outcomes and practical applications. Another advantage rooted in its interpretability is its simplicity; it does not necessitate extensive preprocessing, such as pre-filtering or windowing, nor does it require the level of tuning typically associated with deep learning approaches.

Beyond the development of ConceFT-S, the primary advantage of the proposed analysis framework, ConceFT, lies in its capacity to investigate spindle IFs. The IF of an oscillatory time series, along with its associated phase, encapsulates intricate physiological dynamics. The quantification of these features through TF analysis tools has paved the way for various clinical applications, including automatic sleep apnea detection \cite{lin2016sleep}, early prediction of acute hemorrhage \cite{eid2023using}, and numerous others. In the context of spindle analysis, it's crucial to remember that the alternation between the acceleration and deceleration of sleep spindle frequencies is linked to sleep-related disorders, such as sleep apnea \cite{carvalho2014loss}. Consequently, it becomes an intriguing subject to further investigate whether the more detailed dynamics of spindle IF contain valuable physiological or clinical insights.

The present study has certain limitations, and topens avenues for future exploration. First, it is important to note that our analysis was conducted using a single EEG channel. While we demonstrated its performance in this context, its applicability to different channels remains unvalidated. Furthermore, it is well-established that sleep spindles originating from various brain regions are associated with distinct generators \cite{hahn2019developmental}. Prior research, such as that presented in \cite{dehghani2011topographical}, has shown that spatio-spectral-temporal analysis can illuminate the potential involvement of spindles in coordinating cortical activity during consolidation. The high-resolution TFR generated by ConceFT might be valuable in distinguishing the characteristics of spindles recorded from different channels. Second, our focus has been on epochs labeled as N2 stages by experts, which may limit practical application, given that expert sleep stage annotations may not always be readily available. In such cases, it would be beneficial to employ automatic sleep stage classification algorithms \cite{liu2020diffuse}. Third, although the algorithm has been validated using two publicly available datasets, further validation on larger datasets and application to clinical scenarios is warranted. Lastly, our model \ref{Model:equation} is purely phenomenological and serves the objectives outlined in this paper. However, its potential for further exploration is not guaranteed. One possible direction for enhancement involves incorporating physiological evidence. For instance, there is strong and significant correlation between spectral and temporal features over pre-spindle and spindle periods \cite{gomez2021spectral}, suggesting a connection between the appearance of spindles and the stochastic component $\Phi$. This could be considered in future models.

\section{Conclusion}\label{section conclusion}
ConceFT represents a novel nonlinear-type TF analysis tool that holds significant potential for visualizing and analyzing the dynamics of biomedical signals. The central emphasis of this paper is its application in the study of sleep spindles. Through our research, we have demonstrated that ConceFT enables the development of a straightforward, interpretable, and precise automatic sleep spindle detection algorithm. Additionally, it facilitates the exploration of spindle dynamics, specifically with regards to its IF. The clinical relevance and implications of ConceFT in the brain wave research will be further investigated in our forthcoming work.

\section*{Acknowledgment}
The authors thank Dr. Kraines and the Department of Mathematics at Duke University for funding this project through the PRUV program and offering computing resources. H.-T. Wu thank Dr. Anna Mullins for indicating literature about the sleep spindle acceleration and deceleration.

\bibliographystyle{ieeetr}
\bibliography{Spindle_ConceFT}

\end{document}